\documentclass[journal=enfuem,manuscript=article]{achemso}
\pdfoutput=1
 \usepackage{float}
 \usepackage[labelfont=bf,labelsep=period]{caption}
 \usepackage{sectsty}
 \usepackage{graphicx}
 \usepackage[utf8]{inputenc}
 \usepackage{array}
 \usepackage{geometry}
 \usepackage{setspace}
 \usepackage{xkeyval}
 \usepackage{gensymb}
 \usepackage{xcolor}
 \usepackage{natbib}
 \usepackage{comment}
 \usepackage{soul}
 \usepackage{epstopdf}
 \graphicspath{ {./figs/} }

\usepackage{xr}
\makeatletter
\newcommand*{\addFileDependency}[1]{
	\typeout{(#1)}
	\@addtofilelist{#1}
	\IfFileExists{#1}{}{\typeout{No file #1.}}
}
\makeatother

\newcommand*{\myexternaldocument}[1]{
	\externaldocument{#1}
	\addFileDependency{#1.tex}
	\addFileDependency{#1.aux}
}

\myexternaldocument{supporting}
 
\author{Gilmar F. Arends}
\author{Jae Bem You}
\author{John M. Shaw}
\author{Xuehua Zhang}
\affiliation{Department of Chemical and Materials Engineering, University of Alberta, Alberta T6G 1H9, Canada}
\email{xuehua.zhang@ualberta.ca}

\title{Enhanced displacement of phase separating liquid mixtures in 2D confined spaces}

\begin{document}

\begin{abstract}

Displacing liquid in a confined space is important for technological processes, ranging from porous membrane separation to CO$_{2}$ sequestration. The liquid to be displaced usually consists of multiple components with different solubilities in the displacing liquid. Phase separation and chemical composition gradients in the liquids can influence the displacement rate. In this work, we investigate the effects of liquid composition on the displacement process of ternary liquid mixtures in a quasi-2D microchannel where liquid-liquid phase separation occurs concurrently. We focused on model ternary mixtures containing 1-octanol (a model oil), ethanol (a good solvent) and water (a poor solvent). These mixtures are displaced with water or with ethanol aqueous solution. As a comparison, for some experiments, water was displaced by ternary mixtures. The bright field and fluorescence imaging measurements reveal distinct phase separation behaviours. The spatial distribution of subphases arising from phase separation and the displacement rates of the solution are impacted by the initial ternary solution composition. The boundary between the solution and displacing liquid changes from a defined interface to a diffusive interface as the initial 1-octanol composition in the solution is reduced. The displacement rate also varies non-linearly with the initial 1-octanol composition. The slowest displacement rate arises in the intermediate 1-octanol concentration, where a stable three-zone configuration forms at the boundary. At very low 1-octanol concentration, the displacement rate is fast, associated with droplet formation and motion driven by the chemical concentration gradients formed during phase separation. The excessive energy provided from phase separation may contribute to the enhanced displacement at intermediate to high 1-octanol concentrations, but not at the low 1-octanol concentration with enhancement from induced flow in confinement. The knowledge gained from this study highlights the importance of manipulating phase separation to enhance mass transport in confinement for a wide range of separation processes.

\end{abstract}

\section{Introduction}

Enhanced flow transport in confined spaces is important in many processes, from catalytic-based chemical conversion \cite{Kortunov2005}, porous membrane separation \cite{Tao2018}, nanomedicine \cite{Radha2019, Sanhai2008,Sparreboom2009}, and water treatment \cite{Wu2017}, to enhanced oil recovery \cite{ZhangY2017, Druetta2020} and CO$_2$ sequestration \cite{Gao2017,Feng2019}. In confined spaces, liquid mixing is dominated by slow mutual diffusion \cite{Tartakovsky2019, Frediksen2018}, influenced by the physical properties of liquids. Extensive studies have been performed to understand the effects of wettability of the wall surface \cite{Frediksen2018, Aziz2020, Wu2017}, fluid viscosity \cite{Ho2019, Guillen2012}, or interfacial tension of immiscible phases \cite{Feng2019}. On an industrial scale, the liquid displaced from confined spaces usually consists of multiple chemical constituents that can interact with the displacing liquid \cite{Xu2014, Guillen2012, Zhou2019}. For example, in enhanced oil recovery, chemical flooding processes use a displacing liquid to recover the entrapped oil phase from porous rocks by altering the interfacial tension or by reducing the viscosity of the pore fluid \cite{Xu2014, Mandal2010}. It is important for both fundamental understanding and practical applications to achieve fast fluid movement in confined spaces.

Enhanced fluid transport and autonomous motion of droplets and colloidal particles driven by chemical concentration gradients in multicomponent liquid mixtures \cite{Lohse2020} are topics of current research interest. Microparticles or microdroplets move as a result of diffusiophoretic or solutal driven Marangoni phenomena \cite{Lohse2020}. Concentration gradients are often created and sustained by chemical reaction \cite{Shi2016}, dissolution of surfactants \cite{Nery2017} and surface chemistry interactions \cite{Banarjee2016}. Manipulating the electrolyte concentration in the displacing fluid has been demonstrated to drive colloid transport into and out of microchannels dead-ends \cite{Kar2015, Shin2016, Ault2018}. 

In addition to chemical concentration gradients present in multicomponent single-phase liquid mixtures, subphase formation from phase separation in multicomponent systems is also advantageous in displacing liquid in confined spaces. Even from a practical application perspective, several studies have indicated improvements in oil recovery efficiency from porous media when microdroplet emulsions are formed during displacement compared to cases without emulsion formation. In these cases, the enhancement effect is attributed to the desired rheological properties and seepage characteristics of surfactant-stabilized emulsions \cite{Zhou2019}. More fundamentally, microdroplets form spontaneously from the simple addition of a poor solvent into a ternary solution via the Ouzo effect. When the ternary solution is diluted by a poor solvent that is miscible with the solvent but immiscible with droplet liquid, the mixture becomes oversaturated, leading to the formation of many micro-sized droplets dispersed over the entire continuous liquid phase without the use of mechanical agitation or surfactant \cite{Vitale2003, Lu2017, Zemb2016}. In confinement, the spontaneous formation of microdroplets leads to enhanced transport of the penetrating fluid \cite{Lu2017}. While the effect has been demonstrated, it has not been explored in detail. It remains unclear how the displacement rate is influenced by the relative and absolute compositions of phases formed during phase separation - even for a single mixture.

In this work, we use a simple ternary liquid mixture as a model system to explore the displacement process in quasi-2D confined spaces. More specifically, the confined liquid comprises a ternary solution of 1-octanol, ethanol, and water. The displacing liquid is water. The displacement process is shown to be strongly dependent on the composition of the confined liquid. We demonstrate that over one range of compositions, the confined liquid is displaced. Over another range of compositions, we demonstrate the selective separation and displacement of 1-octanol. As the composition trajectories and hence the intersection points with the one-phase to two-phase boundary in the phase diagram are known, our findings suggest that an effective approach to enhance mass transport in confined spaces is to use displacement liquid composition as a design variable.

\section{Experimental Section}
\subsection{Chemicals and solution preparation}
Chemicals were used as received without further purification. Solution A consisted of 1-octanol (ACS grade $>$95\%, Fischer Scientific), ethanol (Histological grade, Fischer Scientific) and (Milli-Q) water. The compositions of Solution A were varied from 2\% 1-octanol to 50\% 1-octanol as listed in Table \ref{Experimental Conditions Table}. Solution B consisted of (Milli-Q) water. Additional experiments were performed with variations to both confined and displacing liquids, as summarized in the Table \ref{Experimental Conditions Additional Table} below.

The surfaces used for the experiments were hydrophilic glass slides (Fisherbrand Microscope Slides) and hydrophobic silicon wafers. Before use, both substrates were initially washed with piranha solution consisting of 70\%-volume $H_2SO_4$ (ACS Plus grade, Fischer Scientific) and 30\%-volume $H_2O_2$ (30\% ACS grade, Fischer Scientific) at 85$\degree$ C for 20 minutes (caution: piranha solution is highly caustic). They were further cleaned and sonicated with water and ethanol each for 15 minutes. To render the silicon wafers hydrophobic, they were coated with octadecyltrichlorosilane (OTS-Si), as has been previously documented \cite{ZhangX2008}. In brief, 0.5\%-volume of OTS in hexane mixture was used to soak the wafers for 12 hours at room temperature. The OTS-Si substrates were then sonicated with hexane, acetone, and ethanol for 10 minutes each to remove any excessive OTS on the surface. 

For fluorescence experiments, Nile Red (Fischer Scientific) was used due to its high solubility in hydrocarbon-rich phases where the oil-rich domains become fluorescent, which is useful for identifying liquid regions rich in 1-octanol. 

\begin{table}[H]
\captionsetup{font = {small}}
\caption{Evaluated experimental compositions of solution A (displaced liquid) by mass percentage.}
\centering
\begin{tabular}{|c| c| c| c|}
\hline
Composition & 1-Octanol & Ethanol & Water \\
& Mass \% & Mass \% & Mass \%\\
\hline
1 & 50 & 40 & 10 \\
\hline
2 & 40 & 45 & 15 \\
\hline
3 & 35 & 48 & 17 \\
\hline
4 & 30 & 48 & 22 \\
\hline
5 & 25 & 48 & 27 \\
\hline
6 & 20 & 48 & 32 \\
\hline
7 & 15 & 48 & 37 \\
\hline
8 & 10 & 48 & 42 \\
\hline
9 & 8 & 47 & 45 \\
\hline
10 & 5 & 48 & 47 \\
\hline
11 & 2 & 48 & 50 \\
\hline
\end{tabular}
\label{Experimental Conditions Table}
\end{table}

\begin{table}[H]
\captionsetup{font = {small}}
\caption{Compositions of solution A (displaced liquid) and solution B (confined liquid) by mass percentage in additional experiments.}
\begin{tabular}{|c|c|c|c|c|c|c}
\hline
\multicolumn{1}{|l|}{} & \multicolumn{3}{c|}{Solution A} & \multicolumn{3}{c|}{Solution B}   \\ \hline
\multicolumn{1}{|l|}{Composition} &
  \multicolumn{1}{l|}{1-octanol} &
  \multicolumn{1}{l|}{Ethanol} &
  \multicolumn{1}{l|}{Water} &
  \multicolumn{1}{l|}{1-octanol} &
  \multicolumn{1}{l|}{Ethanol} &
  \multicolumn{1}{l|}{Water} \\ \hline
E1                     & 50       & 40       & 10        & 0  & 25 & \multicolumn{1}{c|}{75} \\ \hline
E4                     & 10       & 48       & 42        & 0  & 25 & \multicolumn{1}{c|}{75} \\ \hline
R1                     & 0        & 0        & 100       & 50 & 40 & \multicolumn{1}{c|}{10} \\ \hline
R4                     & 0        & 0        & 100       & 10 & 48 & \multicolumn{1}{c|}{42} \\ \hline
\end{tabular}
\label{Experimental Conditions Additional Table}
\end{table}

\subsection{Dimensions and setup of the microchamber}
\begin{figure}[H]
	\includegraphics[trim={0cm 0cm 0cm 0cm}, clip, width=0.6\columnwidth]{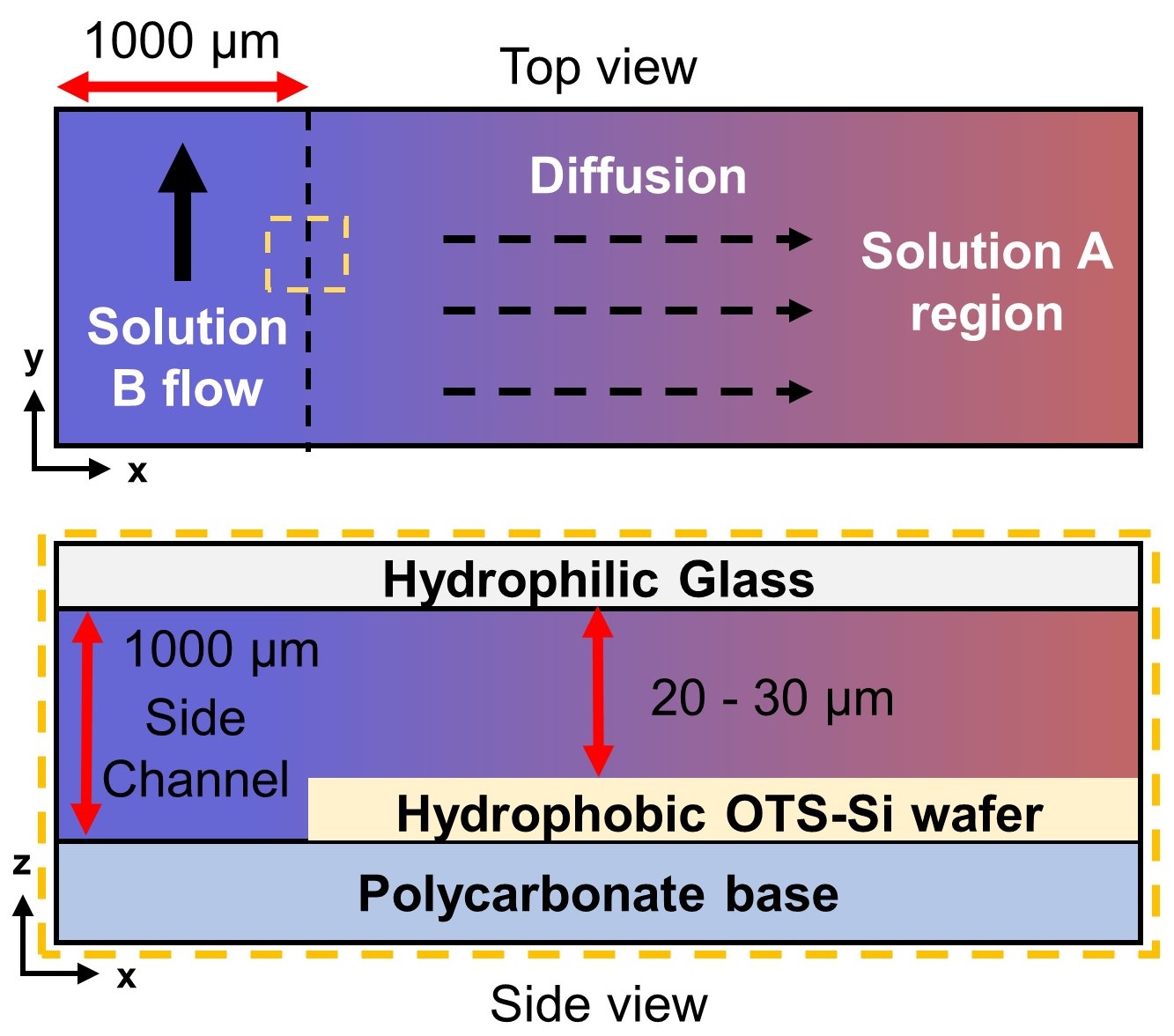}
	\caption{Sketch of the fluid microchamber with top view and cross-section of the main and side channels.
	}
	\label{experimentalsketch}
\end{figure}

The flow chamber, as sketched in Figure \ref{experimentalsketch}, consisted of a polycarbonate base (8.5 $\times$ 4.5 $\times$ 1 cm), with the hydrophilic cover glass as the top sealed with a silicon rubber spacer. Before any experimentation, every component from the chamber was sonicated in both water and ethanol for 10 minutes each. The chamber was assembled by connecting the tubing to the polycarbonate base as the inlet and outlet streams. A droplet of Solution A was placed on the OTS-Si substrate. The substrate was held by a capillary bridge to the cover glass. A small channel height of around 20 to 30 $\mu m$ was kept. The cover glass was then held in position by a spacer and was clamped to the base of the chamber with large binder clips.

\subsection{Process of the liquid displacement}

Solution B was pumped using a syringe pump (Fisherbrand Single Syringe Pump) to the channel. The displacement process as water diffused transversely into the narrow channel was captured in-situ using a camera (Nikon DS-Fi3) connected to an optical microscope (Nikon Eclipse Ni-U). The magnification under normal conditions was 10$\times$ with a F.O.V. of 22 $mm$, and videos were captured using an auto-exposure setting leading to captured footage with framerates ranging from 10-15 frames per second (FPS) at a fixed resolution of 2880 $\times$ 2048. Fluorescence footage required longer exposure times to capture the fluorescent features leading to low frame rates of 0.9 to 1 FPS.

\subsection{Analysis of the displacement rates}

The videos were processed through ImageJ by converting them to image sequences. If required, the contrast and brightness levels were adjusted, and a bleach correction was applied. The tracking of the boundaries was achieved by the manual tracking plugin within the program. The calibration used for the captured footage at 10x magnification was 0.24 px/$\mu$m. The tracked data were saved as a raw CSV file, including the number of frames, x and y positions in pixels. The displacement rate was calculated using the x and y position using the first entry in the data set as the initial position.

\section{Results and Discussion}

\begin{figure}[H]
	\includegraphics[trim={0cm 0cm 0cm 0cm}, clip, width=1\columnwidth]{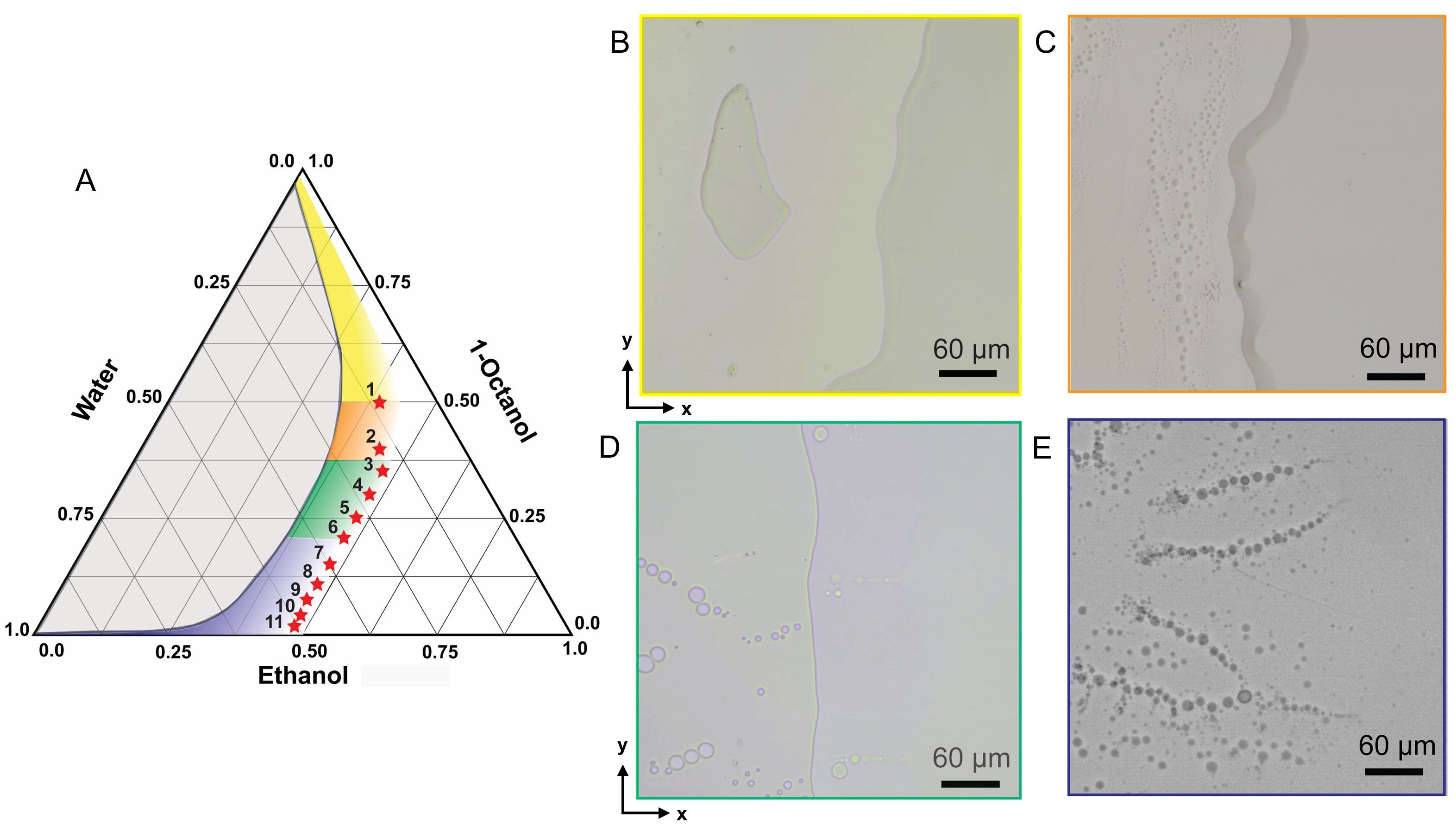}
	\caption{Overview of different dynamic regimes. (A) Ternary phase diagram of 1-octanol, ethanol, and water system in terms of mass fractions with evaluated experimental compositions annotated. The phase diagram is adapted from ref \cite{Lopian2016}. Displacement regimes are indicated by colors: Regime 1 (yellow), Regime 2 (orange), Regime 3 (green), and Regime 4 (blue). (B-E) Representative optical images of the phase separation regimes are shown in colour-coded rectangles as in (A).
	}
	\label{overview}
\end{figure}

The phase diagram for the model ternary mixture 1-octanol (oil), ethanol (good solvent) and water (poor solvent) is illustrated in Figure \ref{overview}A. Confined liquid compositions labelled from 1 to 11 (Table \ref{Experimental Conditions Table}), are shown in the phase diagram. The phase boundary was adapted from the literature \cite{Lopian2016}, tie lines not shown were computed using the UNIFAC model and inform the discussion. The Dortmund UNIFAC model was implemented using the parameter values noted in the Supporting Information (Tables S1 $\sim$ S4).

The mixing between solution A and water as water diffused into the narrow channel induced liquid-liquid phase separation. Below we show the displacement process along with phase separation for the wide range of solution A compositions listed in Table \ref{Experimental Conditions Table}. The 1-octanol-rich liquid is revealed using a hydrophobic dye in fluorescence images. The displacement dynamics are categorized into four regimes, as shown in Figure \ref{overview}B-E. 

\subsection{Regime 1: Receding interface}

\begin{figure}[H]
\includegraphics[trim={0cm 0cm 0cm 0cm}, clip, width=0.80\columnwidth]{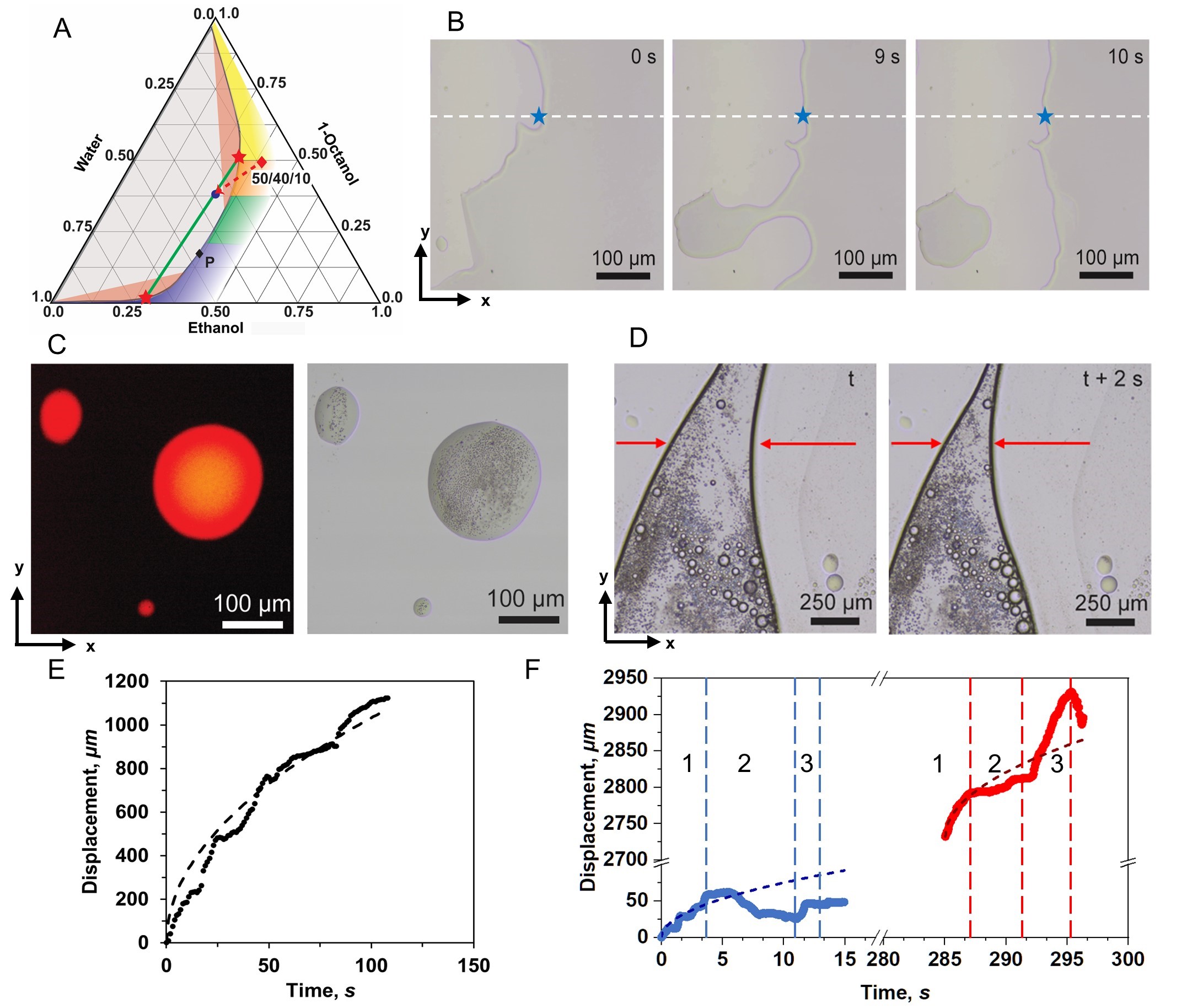}
	\caption{Overview of Regime 1. (A) Ternary phase diagram indicating the dilution path with the two macroscopic subphases formed. P is the plait point of the ternary mixture. Red regions within the phase envelope are the Ouzo and reverse Ouzo regions. Adapted from ref \cite{Lopian2016}. The initial solution A composition is given by a red diamond. An illustrative unstable mixture composition during dilution is given as a blue diamond. The red stars represent the 1-octanol-rich and water-rich compositions formed along the green tie line. (B) Time-based optical images of early film and domain development. (C) Side by side fluorescence and bright-field images of representative microdomains. (D) Time-based images of a deformed domain narrowing with time. (E) Plot showing local boundary displacement without microdomain formation. (F) Plot showing local boundary displacement at a location where microdomains form. Blue data set was taken along the x-direction indicated by white dashed line in (B) with the same reference position of the side channel. Time window 1 is for a period without a microdomain present. During time period 2, a microdomain forms at the boundary. Time period 3 starts as soon as the microdomain detaches from the boundary. The formation of microdomains slows boundary displacement. Their detachment accelerates boundary displacement.}
	\label{Regime 1}
\end{figure}

Regime 1 corresponds to a high 1-octanol concentration, above 50\% by mass in solution A. The composition is labelled in the phase diagram in Figure \ref{Regime 1}A. As water diffused from the side channel, the resulting images in Figure \ref{Regime 1}B show a clear interface separating solution A from water in the narrow channel.  

A few 1-octanol microdomains pinched off from the receding phase boundary at several locations and remained attached to the hydrophobic wall on the water side of the boundary. Their composition was confirmed using fluorescence imaging (Figure \ref{Regime 1}C). This fluorescence image of a representative domain formed from solution A doped with an oil-soluble fluorescent dye (Nile red). The high fluorescence intensity from the microdomain indicates the high concentration of the dye in the microdomain, in contrast to the absence of a fluorescence signal from the surrounding liquid. The spatial distribution of the dye is consistent with the chemical composition of the 1-octanol-rich microdomains surrounded by water-rich liquid. 

In this regime, water diffusing into the narrow channel did not lead to liquid-liquid phase separation along the boundary - a behaviour consistent with that of an immiscible displacing liquid. However, liquid-liquid phase separation did occur within the stranded microdomains. Many nanodroplets (with a diameter of approximately 4 $\mu m$) formed, as shown in Figure \ref{Regime 1}D. The formation of these water-rich nanodroplets is attributed to subphase formation arising from diffusion of water into the microdomains followed by liquid-liquid phase separation. 

As water diffuses into a microdomain and 1-octanol and ethanol diffuse out, the global composition of the microdomain changes. To a first approximation, the dilution path that is followed is indicated by the dashed line in the phase diagram in Figure \ref{Regime 1}A. The mixture becomes unstable as the composition intersects the solubility curve and spontaneously forms water-rich and 1-octanol-rich subphases. Although the global composition of the microdomain at separation is unknown, the compositions of the 1-octanol-rich and water-rich subphases created are approximately 80 and 19\% 1-octanol by mass. It is interesting that phase separation occurred in the microdomains of stranded solution A, but not at the boundary between water and solution A. The large surface to volume ratio of the small microdomains permits them to become supersaturated with water quickly. By contrast, at the boundary, water diffuses into the bulk 1-octanol-rich phase, and 1-octanol and ethanol diffuse into the bulk water-rich phase. Consequently, the liquids on both sides of the boundary do not supersaturate within the time frame of measurements. At longer times, one would certainly expect the microdomains to dissolve into the water-rich phase.

While the shape and size of stranded solution A microdomains varied between experiments (Figure \ref{Regime 1}D), possibly due to the nature of the interface instability - subphase formation within them over time was common.

If we now focus on the displacement of the boundary between solution A and water, in the absence of microdomain formation, we can see that the x-location displacement rate is essentially time-invariant with an average value of 10.7 $\pm$ 0.3 $\mu m/s$ (Figure \ref{Regime 1}E). Figure \ref{Regime 1}F shows the fluctuating x-location displacement rate where microdomains form and detach over time. The interface progressively displaced solution A in the positive x-direction. During the formation of a microdomain, the interface remains stationary or moves backward. When a microdomain detaches from the interface, the interface accelerates in the positive x-direction. Such cycles were observed at multiple locations on the surface.

The blue and red dashed curves in Figure \ref{Regime 1}F are 1D diffusion-based displacement curves, $l=(2Dt)^{1/2}$, fitted to the interface displacement data before microdomain detachment to obtain an effective diffusion constant \cite{Lu2017}. The effective diffusivity varies as the boundary progresses. The values range from 5.3 $\times 10^{-9}$ $m^2/s$ for a part of the boundary without microdomain formation (Figure \ref{Regime 1}E) and 2.8 $\times 10^{-10}$ to 7.9 $\times 10^{-10}$ $m^2/s$ for a location where there is microdomain formation. The mutual diffusivities of water and ethanol, and water and octanol range from 1.08 $\times 10^{-10}$ to 1.23 $\times 10^{-9}$ $m^2/s$ and 2.0 $\times 10^{-10}$ to 7.3 $\times 10^{-10}$ $m^2/s$, respectively \cite{Pratt1974, Kinoshita2017, Su2010}. So the diffusion rates are consistent with expectations. In locations with microdomain formation, the boundary displacement rate was variable and ranged from 4.3 to 14.0 $\mu m/s$.

The influence of solution A microdomains on the local motion of the boundary may be attributed to the pinning effect and mass transfer to microdomains. Imperfections on the wall surface pin the boundary, leading to microdomain formation once the boundary becomes unstable. As the microdomain is formed, the water-rich phase close to the microdomain experiences interfacial tension opposite to its flow direction. The acceleration experienced after the detachment of the microdomain may be attributed to the release of built-up tension that comes from the stretched interface of the microdomain.

In Regime 1, water largely displaces solution A (with a high initial concentration of 1-octanol) in a confined space with one hydrophobic wall. The large scale stability of the boundary between solution A and water contributes to the effective displacement. In contrast, local instability of the boundary and microdomain formation leads to lower and fluctuating displacement rates indicated in Figure \ref{Regime 1}F, as well as less efficient 1-octanol-rich liquid displacement. 

\subsection{Regime 2: Moving microdroplet zone}

\begin{figure}[H]
\includegraphics[trim={0cm 0cm 0cm 0cm}, clip, width=0.9\columnwidth]{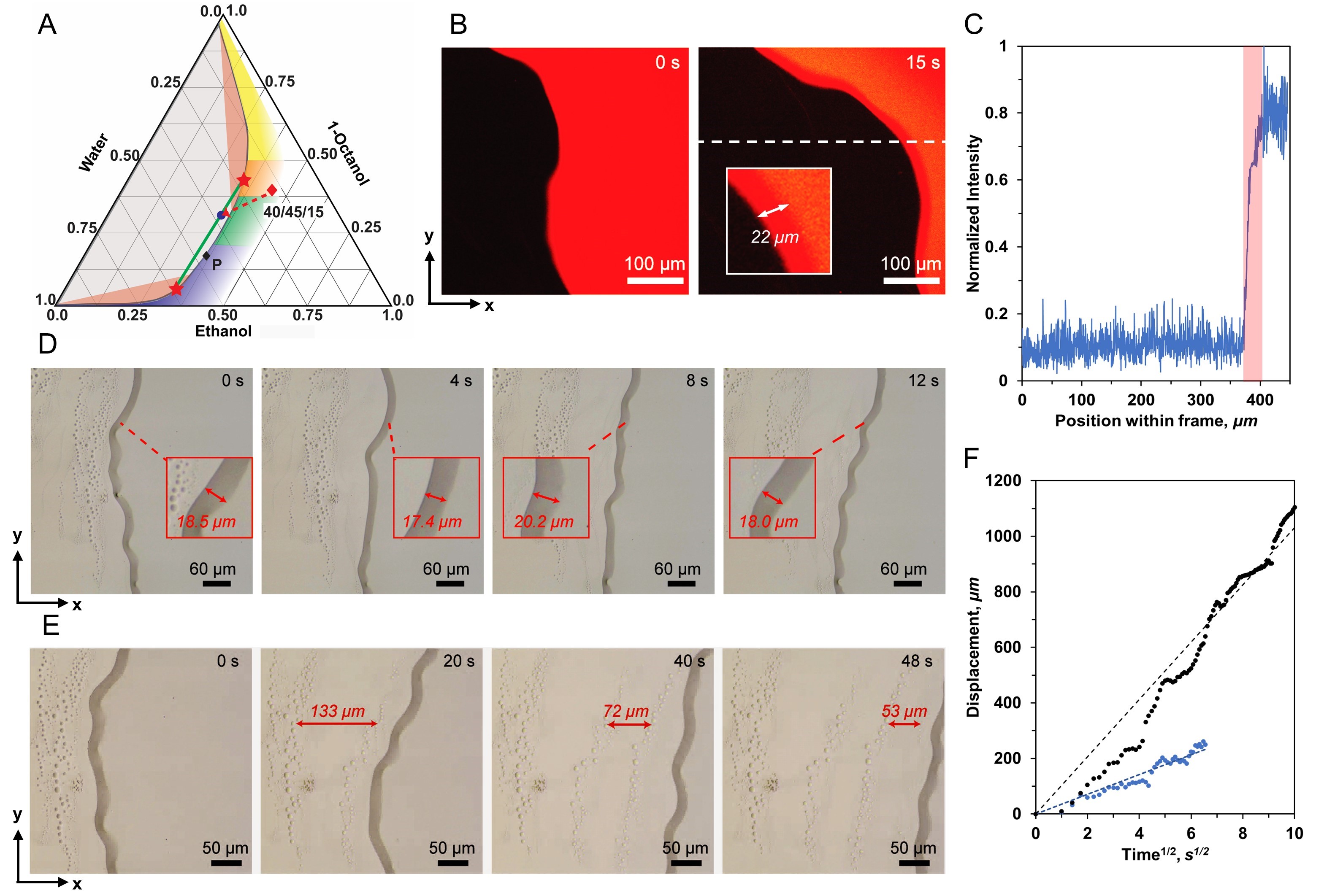}
	\caption{Overview of Regime 2. (A) Ternary phase diagram indicating the approximate dilution path and the compositions of the subphases formed. P is the plait point of the ternary mixture. Red regions within the phase envelope are the Ouzo and reverse Ouzo regions. Adapted from ref \cite{Lopian2016}. The initial solution composition is given by a red diamond. The unstable mixture composition after dilution is approximated by a blue diamond. The red stars represent the 1-octanol-rich and water-rich subphases that are formed based on the green tie line. (B) Time-based fluorescence imagery at 40\% 1-octanol within Regime 2. (C) Normalized fluorescence intensity as a function of x-location within the frame of (B) at t = 15 s. The red area in the graph indicates a boundary zone. (D) Time-based optical images of moving boundary within Regime 2 with an initial solution A composition of 40\% 1-octanol. Inset contains magnified images of the moving boundary indicating zone thickness. (E) Time-based optical images of droplet formation in the boundary zone. (F) Boundary displacement as a function of the square root of time assuming diffusion-based dynamics. Regime 1 smooth interface displacement (black) and Regime 2 (blue) comparison.
	}
	\label{Regime 2}
\end{figure}

Regime 2 corresponds to intermediate 1-octanol concentrations, about 40\% by mass in solution A. The overall behaviour in Regime 2 differs markedly from Regime 1, despite the similarity of the solution A compositions. Starting from this composition, the subphase formed by mixing with water is well approximated by the trajectory shown in Figure \ref{Regime 2}A. From the tie line shown in the phase diagram, much more 1-octanol-rich subphase is still expected than water-rich subphase following liquid-liquid phase separation. Unlike Regime 1, a mobile 1-octanol-rich microdroplet zone separates solution A and water throughout the narrow channel, as shown in the fluorescence images in Figure \ref{Regime 2}B. A dark zone containing water, a zone rich in microdroplets and a clear zone of solution A are clearly present.

Figure \ref{Regime 2}C shows the level of fluorescence intensity crossing the three zones. The high fluorescence intensity in the intermediate zone confirms that it is rich in 1-octanol. The boundary zone appears dark in bright field images, possibly due to scattering from small 1-octanol-rich drops.

The time-based images in Figure \ref{Regime 2}D show oscillation of the boundary zone, but this oscillation does not lead to the formation of stranded microdomains. Instead, the boundary zone thickness appears to remain constant at 18-20 ${\mu}m$ in the time interval of 12 $s$ as the boundary moves forward, as shown in the high magnification inset (Figure \ref{Regime 2}D). The time invariance of the boundary zone thickness may indicate that the mass transfer occurring through the boundary is balanced, even though the compositions of the water-rich and 1-octanol-rich phases in the zone vary along the x-direction.

Lines of stranded 1-octanol-rich droplets form behind the receding boundary zone, mirroring the shape of the boundary as it progresses (Figure \ref{Regime 2}E). The interval between two lines of droplets on the surface may be related to the time required for phase separation to occur in the boundary zone or the time required for the droplets inside the boundary zone to accumulate on the contact line with water. 

The rate of the boundary displacement is quantified in Figure \ref{Regime 2}F along a fixed-line in the x-direction. The rate of displacement fluctuates with time due to the oscillatory advancement of the boundary zone. Again, the data can be fitted with a diffusion model \cite{Lu2017}. Compared to the displacement of the boundary in Regime 1 (for a case where microdomain formation does not occur), the displacement versus square root of time is slower in Regime 2. This is evidenced by the lower effective diffusivity obtained from the fit, which ranged from 5.8 $\times 10^{-10}$ to 6.7 $\times 10^{-10}$ $m^2/s$, and a lower displacement rate of 5.8 $\mu m/s$.

\newpage
\subsection{Regime 3: Moving three-zone configuration}

\begin{figure}[H]
	\includegraphics[trim={0cm 0cm 0cm 0cm}, clip, width=1\columnwidth]{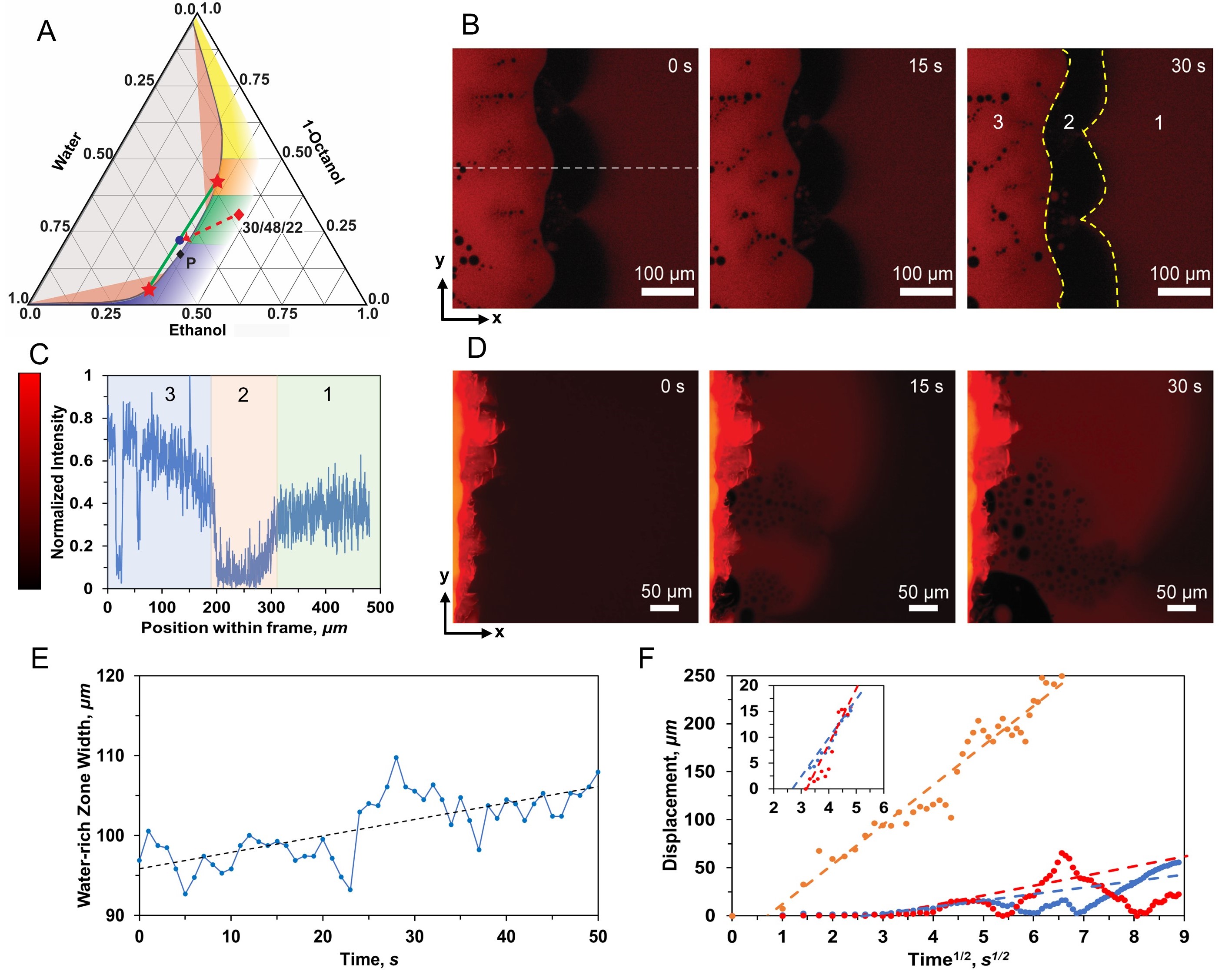}
	\caption{Overview of Regime 3 and dynamic data for 30\% 1-octanol in solution A. (A) Ternary phase diagram showing the approximate dilution path and subphases compositions. P is the plait point of the ternary mixture. (B) Illustrative time-based fluorescence images for Regime 3. Zone 1 is a 1-octanol-rich phase, zone 2 is a water-rich phase, and zone 3 is 1-octanol-rich liquid with dispersed water-rich phase drops. (C) Normalized fluorescence intensity as a function of x-location within the frame of (B) at t = 0 s. (D) Time-based fluorescence images stages as water begins to enter the narrow channel. (E) Water-rich zone 2 width as a function of time (dashed line indicating slight increasing trend). (F) Boundary displacement as a function of the square root of time illustrating diffusion-dominated movement. (Blue) corresponds to the boundary between zones 1 and 2. (Red) corresponds to boundary between zones 2 and 3. (Orange) corresponds to Regime 2 displacement data. Inset depicts the slopes, from 2 $<$ $t^{1/2}$ $<$ 6 for the boundaries in Regime 3.}
	\label{Regime 3}
\end{figure}

Regime 3 arises for the 1-octanol concentration range 20\% to 30\% by mass in solution A. Data for the 30\% case are presented in detail and are representative of this regime. Figure \ref{Regime 3}A illustrates the dilution process within the phase diagram. On separation, approximately equal amounts of water-rich and 1-octanol-rich subphases form because the dilution line intersects the phase envelope close to the plait point.

Fluorescence images in Figure \ref{Regime 3}B reveal three distinct zones along the x-direction as labelled in Figure \ref{Regime 3}B at t = 30 s. The normalized fluorescence intensity plotted in Figure \ref{Regime 3}C shows that zone 1 consists of a 1-octanol-rich phase revealed by partition of a dye \cite{Li2018}, that zone 2 consists of a water-rich subphase, and that zone 3 contains numerous immiscible water-rich droplets (dark features in fluorescence images within a 1-octanol-rich phase).

Figure \ref{Regime 3}D illustrates the early development of the three-zone configuration as water enters from the side to the narrow channel. A 1-octanol-rich phase forms from phase separation, and water droplets appear simultaneously within the 1-octanol-rich subphase and later coalesce and merge into a water-rich zone 2. 

The displacement of solution A proceeds as the entire zone 2 moves into the narrow channel. The width of zone 2 appears to increase slightly with time at a fixed y-position, as shown in Figure \ref{Regime 3}E. The most noticeable aspect of the entire zone 2 is its stability in space over minute long time scales. We suspect that this stability is related to the near-constant width of zone 2, which we in turn attribute to the near equivalence of the influx and outflux of liquid to zone 2 over time. Liquid may enter zone 2 in the form of water-rich droplets from zone 3 or by water entering from the side channel, as evidenced in the Figure S2. Concurrently the 1-octanol-rich liquid leaves zone 2 to zone 3. The balancing of mass fluxes in and out of zone 2 accounts for its stability.

The displacement of the zone 1 to zone 2 and zone 2 to zone 3 boundaries versus the square root of time are shown in Figure \ref{Regime 3}F. These boundaries move with an oscillatory behaviour in the positive x-direction with a slower overall rate than the previous regime 2 at 0.71 $\mu m/s$. The fluctuations were found to be related to the water-rich droplets entering zone 2. At locations were droplets merged with the boundary between zone 2 and 3, the boundary retracted in the negative x-direction. The boundary successively becomes deformed, leading to the acceleration of the boundary in the positive x-direction as the shape of the boundary starts to recover. The boundary displacement curves from the two locations in Figure \ref{Regime 3}F show similar trends in pinning and depinning transitions. Regardless of all the fluctuations in boundary motion, the three-zone configuration remained stable with time due to water replenishment.

\newpage
\subsection{Regime 4: Diffusive boundary}
\begin{figure}[H]
	\includegraphics[trim={0cm 0cm 0cm 0cm}, clip, width=1\columnwidth]{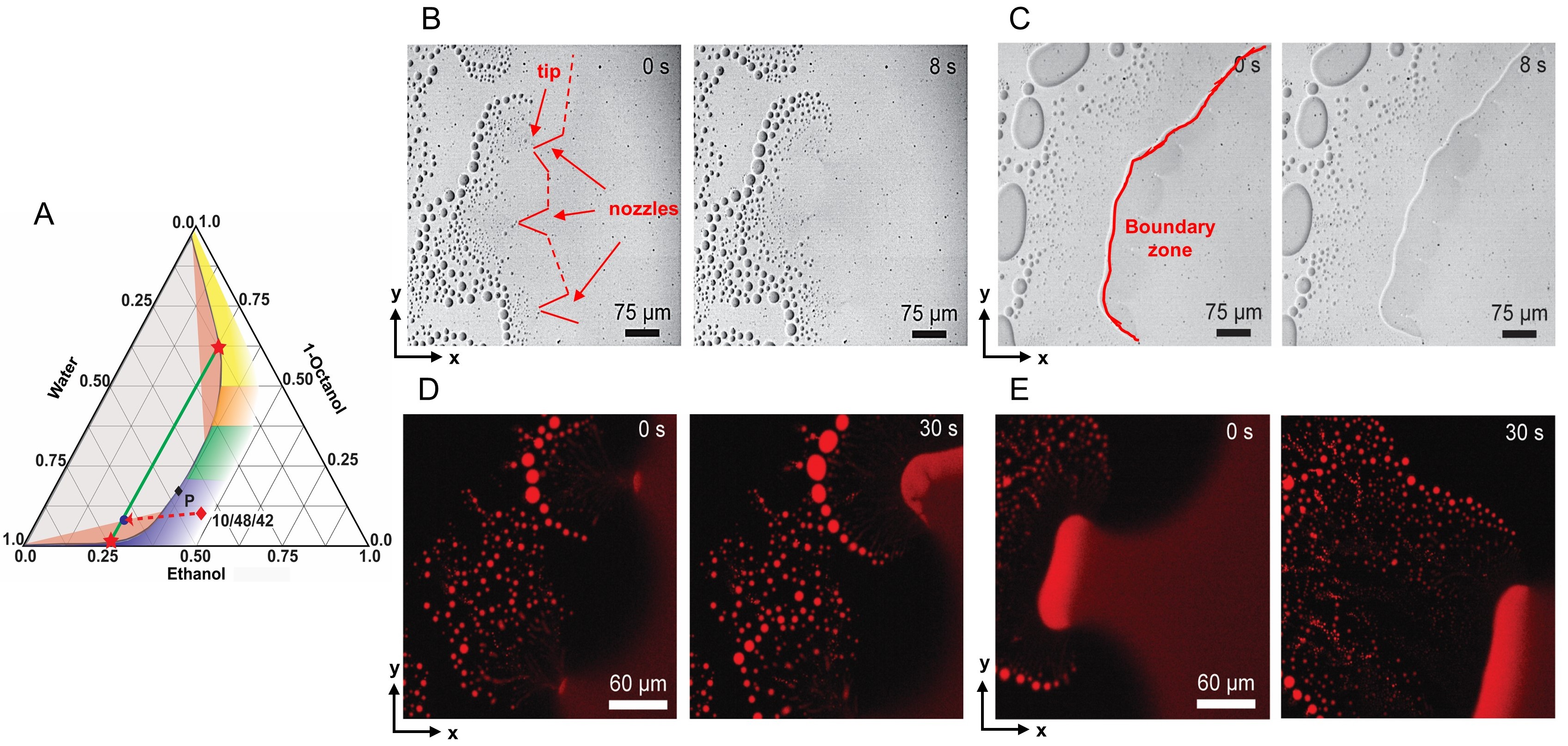}
	\caption{Overview of Regime 4 at low 1-octanol compositions. (A) Ternary phase diagram indicating the approximate dilution path and subphase compositions. P is the plait point of the ternary mixture. (B) Time-based optical images showing boundary development and (C) boundary for the 10\% 1-octanol solution A case. (D) Time-based fluorescence imagery of protrusion and change to (E) line-shaped boundary for the 10\% 1-octanol case.}
	\label{Regime 4 part 1}
\end{figure}

In Regime 4, the composition ranges from 2-20\% 1-octanol by mass in solution A. In this regime, a diffusive boundary forms and the dynamics differ from the other three regimes. The results depicted in Figure \ref{Regime 4 part 1} are for 1-octanol for 10\% by mass.

Figure \ref{Regime 4 part 1}A shows the approximate dilution path and subphase compositions in the phase diagram for the 10\% 1-octanol case. The diluting mixture intersects the phase envelope below the plait point, and following phase separation, the water-rich subphase fraction is expected to an order of magnitude greater than the 1-octanol-rich subphase mass fraction.

Figure \ref{Regime 4 part 1}B shows the diffusive boundary at an early transition stage consisting of numerous triangular protrusions of solution A that extend into the water phase. Droplets formed from the tips of these protrusions in the water-rich phase were larger than those formed at other positions along the boundary because, at the tip of the protrusions, the concentration gradients are sharper than at other positions \cite{Lu2017, Vitale2003}. At longer times, protrusions approach one another and collapse into a line-shaped boundary region, as shown in Figure \ref{Regime 4 part 1}C At longer times, droplets are formed along the entire boundary in the water-rich phase.

Fluorescence images in Figures \ref{Regime 4 part 1}(D, E) show the 1-octanol-rich protrusions with a 1-octanol concentration greater than in solution A. Ethanol transfers prefers preferentially to the water-rich phase. After 30 seconds, the composition of the protrusions becomes consistent with that of the 1-octanol rich droplets formed, as indicated by the fluorescence intensity.

\begin{figure}[H]
	\includegraphics[trim={0cm 0cm 0cm 0cm}, clip, width=0.85\columnwidth]{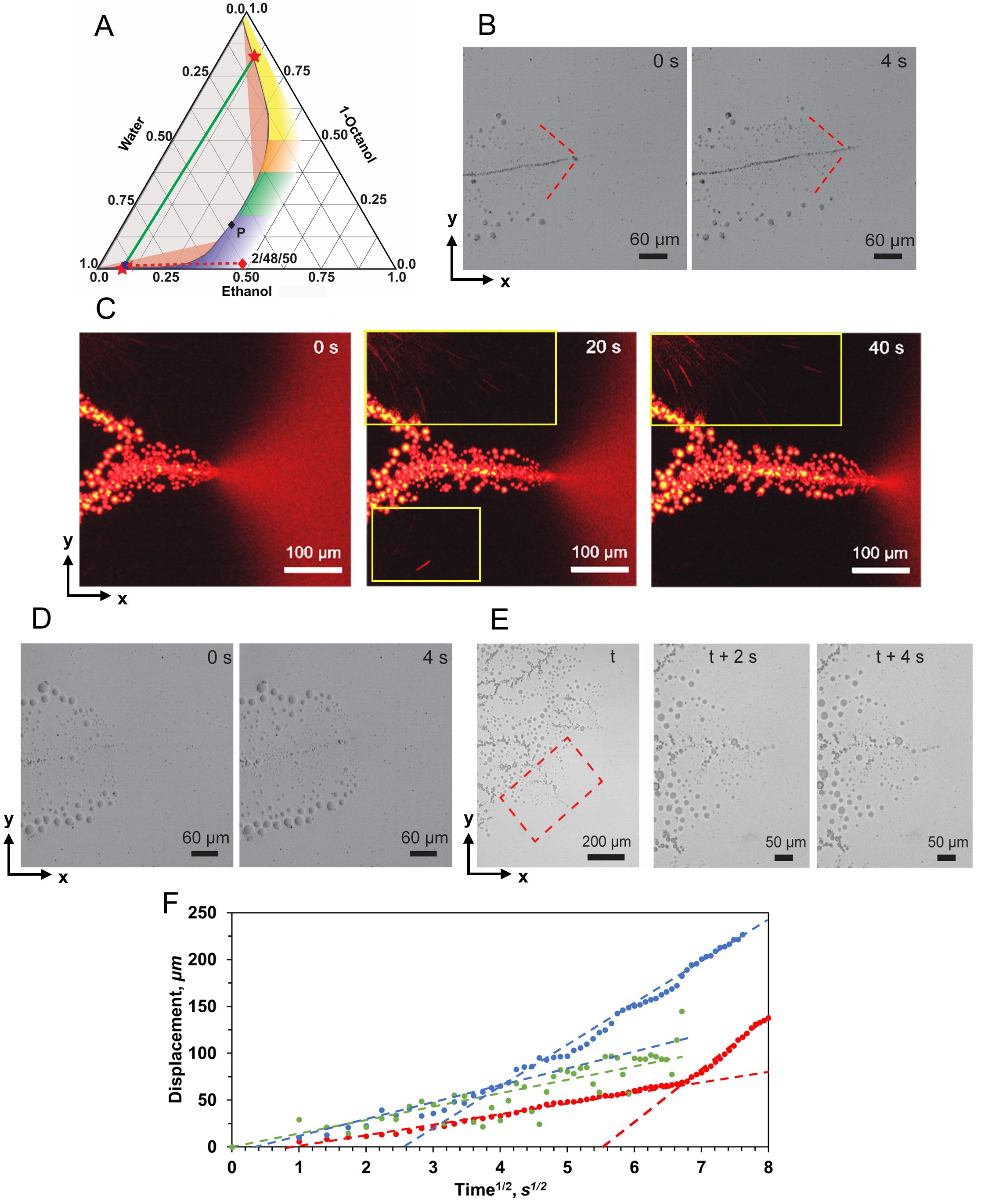}
	\caption{Regime 4 dynamics with a focus on low 1-octanol wt\%. (A) Ternary phase diagram indicating approximate the dilution path and subphases compositions for solution A with an initial 1-octanol composition of 2\% by mass. (B) Time-based optical images of triangular shaped boundary movement using 2\% by mass in solution A. (C) Time-based fluorescence images of 2\% 1-octanol by mass indicating the 1-octanol-rich and water-rich regions. Yellow boxes show droplets recirculating back to the boundary region. Time-based optical images comparing the 5\% 1-octanol (D) and 8\% 1-octanol (E) conditions. (F) Boundary displacement as a function of the square root of time plot comparing the 2\% octanol (blue), 5\% (green) and 10\% (red) conditions. Dashed lines show anticipated two-tiered diffusion-based displacement dynamics.}
	\label{Regime 4 part 2}
\end{figure}

Figure \ref{Regime 4 part 2}A provides an overview and dynamic data for solution A with 2\% 1-octanol. Figure \ref{Regime 4 part 2}A shows the approximate dilution path and subphase compositions. The mass ratio of water-rich to 1-octanol-rich subphase is expected to be around 25 to 1 in this case. Fluorescence images in Figure \ref{Regime 4 part 2}C again show triangular protrusions of 1-octanol-rich liquid into the boundary. While the outcomes are similar to those described in Figure \ref{Regime 4 part 1}, the triangular protrusions are more stable (Figure \ref{Regime 4 part 2}B). This outcome is consistent with earlier work showing that droplets form branches from a diffusion-limited nucleation and growth process in a 2D confinement at low concentration \cite{Lu2015}.

Once formed, numerous droplets travelled toward the side channels in the negative x-direction before becoming immobilized on the wall surface or coalescing with other droplets. The lines with low fluorescent intensity indicated in the yellow boxes in Figure \ref{Regime 4 part 2}C show induced droplet flow. The droplets appear as lines due to the exposure time of the camera when taking fluorescence imagery. This induced flow may explain the relatively fast boundary displacement rate associated with Regime 4.  

The optical images in Figures \ref{Regime 4 part 2}D and E indicate the qualitative similarity of droplet formation and boundary movement phenomena for 5\% and 8\% 1-octanol by mass in solution A. Differences arise from the number of droplets formed and droplet size with time. 

The displacement of the boundary is plotted against the square root of time for the 2\%, 5\% and 10\% initial 1-octanol wt-\% cases. As indicated by the slopes added to each curve in Figure \ref{Regime 4 part 2}F, a slower followed by a more rapid diffusive displacement process is clearly evident for the 2\% and 10\% conditions. While qualitatively similar, the first diffusion-controlled process slows and has a longer duration as the initial 1-octanol wt-\% increases. In Regime 4, boundary displacement is faster than Regime 3 compositions, and there are seldom oscillations in displacement versus time plots.

\subsection{Comparison of the displacement rates in the four regimes}

\begin{figure}[H]
	\includegraphics[trim={0cm 0cm 0cm 0cm}, clip, width=0.90\columnwidth]{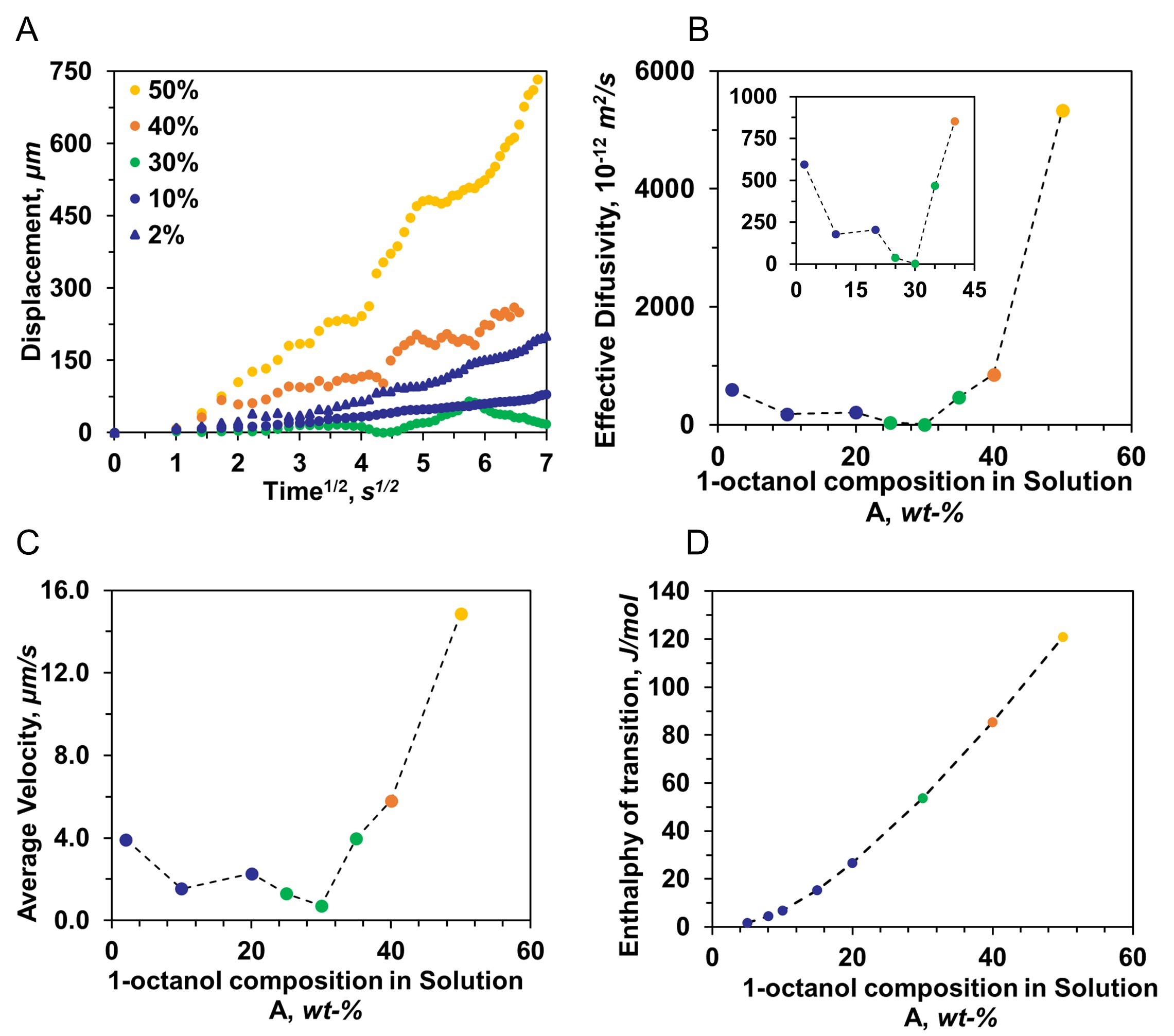}
	\caption{(A) Summary of displacement versus square root of time for 1-octanol compositions spanning the four regimes. (B) Effective diffusivity and average velocity (C) calculated from displacement versus time data in all four regimes. (D) Calculated enthalpy of transition from a non-equilibrium single phase state to two equilibrium states based on the Dortmund UNIFAC method.}
	\label{Summary Displacement}
\end{figure}

In Figure \ref{Summary Displacement}A, we compare the displacement rates in all four regimes presented. An effective diffusivity was calculated for each regime, as shown in Figure \ref{Summary Displacement}B. The highest effective diffusivity value was obtained in Regime 1 at 50\% of 1-octanol for a smooth boundary, which has an enhanced effective diffusivity of $8 \times 10^{-9}$ $m^2/s$. The effective diffusivity obtained is thus higher than the expected range of the mutual diffusivities between water and ethanol (1.1 $\times 10^{-10}$ to 1.2 $\times 10^{-9}$ $m^2/s$), and water and 1-octanol (2.0 $\times 10^{-10}$ to 7.3 $\times 10^{-10}$ $m^2/s$) showing the enhanced diffusion. In both high and low 1-octanol ratios of Regimes 1 and 4, the displacement rate of the boundary was faster at $\sim$ 15 ${\mu}m/s$ and $\sim$ 4 ${\mu}m/s$. In contrast, in the intermediate Regime 3, the rate is only $\sim$ 0.7 ${\mu}m/s$ as shown by the average velocities in Figure \ref{Summary Displacement}C. Boundary displacement in Regime 3 is slower than in the other regimes but remains 2 times the displacement rate expected for a purely-diffusive single-phase process obtained from the effective diffusivity.

To understand the dependence of displacement rate on 1-octanol concentration, including a shallow minimum at intermediate 1-octanol concentrations, we must consider competing effects. Differences in the energy released during phase separation with 1-octanol concentration are illustrated in Figure \ref{Summary Displacement}D. The energy released during the transition from a non-equilibrium single-phase state to an equilibrium two-phase state increases with 1-octanol concentration. To the extent that displacement rate depends on this thermodynamic driving force, one would anticipate an increase in displacement rate with 1-octanol concentration. This effect is clearly evident at high 1-octanol concentration. However, the displacement rate at low 1-octanol concentrations is somewhat higher than at intermediate 1-octanol concentrations. We attribute this apparent enhancement in displacement at low 1-octanol concentrations to droplet propulsion and the flow induced in the continuous phase from the self-propelling microdroplets in this concentration range, i.e., to how the energy of phase transition is dissipated. 

Impacts of relative viscosity of the liquid phases, and the relative wettability of the liquid phases on the wall surface, which approach one at intermediate 1-octanol concentrations (near the plait point), are seen as secondary. Viewed from this perspective, the strong dependence of displacement rate on the initial composition is expected to be a general feature for displacing multi-component liquids undergoing phase separation. Further, compositions transitioning through Regimes 1 and 2 are preferred for displacing liquid, while compositions transitioning through Regimes 3 and 4 are preferred for promoting the separation and extraction of a component or a component category (1-octanol a model oil in this case) from the solution at lower overall displacement rates.

\newpage

\subsection{Using water + ethanol binary mixtures to displace solution A}

\begin{figure}[H]
	\includegraphics[trim={0cm 0cm 0cm 0cm}, clip, width=0.8\columnwidth]{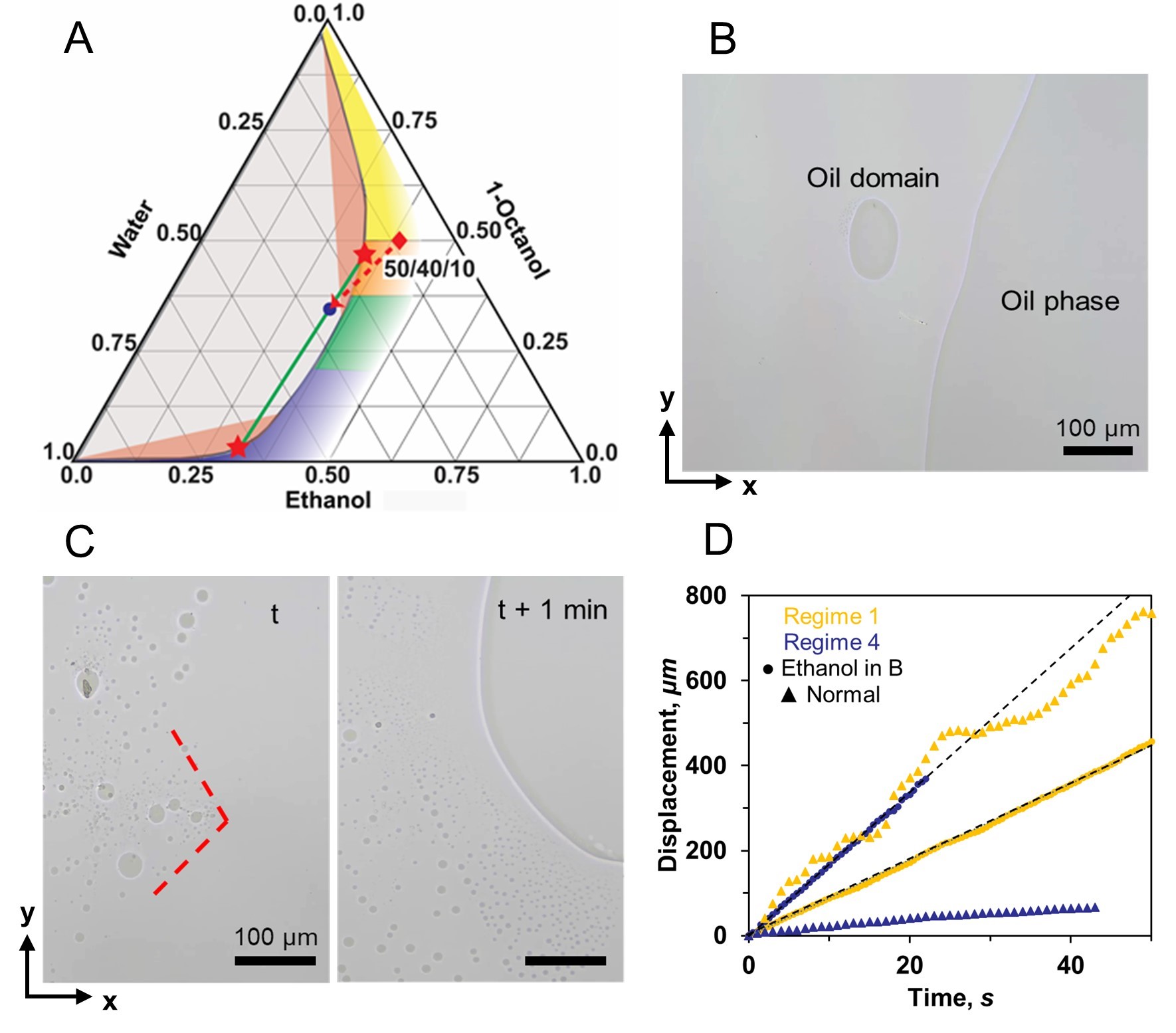}
	\caption{(A) Ternary phase diagram indicating the approximate dilution path using 25\% ethanol in the displacing liquid at Regime 1 condition. (B) Optical image of Regime 1 condition using 25\% ethanol by mass in the displacing liquid. (C) Time-based optical images of Regime 4 condition using 25\% ethanol by mass in the displacing liquid. (D) Boundary displacement as a function time for ethanol in displacing liquid comparing the 10\% 1-octanol (blue) and 50\% (yellow) conditions. Dashed lines indicate linear regression from data. Data for pure water as displacing liquid in both regimes was added for comparison (triangles).}
	\label{Variations to displacing liquid}
\end{figure}

Additional experiments were performed at high and low 1-octanol composition regime conditions using a displacing liquid containing 25\% ethanol by mass in water. Figure \ref{Variations to displacing liquid}A shows the approximate trajectory and associated tie line from the dilution process with this displacing liquid for a Regime 1 composition when water alone is used to displace it (see E1 in Table \ref{Experimental Conditions Additional Table}). The resulting water-rich phase formed by phase separation, in this case, has a composition corresponding to a Regime 4 solution composition, while the composition of the 1-octanol-rich phase remains qualitatively similar.

Figure \ref{Variations to displacing liquid}B shows an observed insoluble moving boundary and stranded 1-octanol-rich microdomains similar to the Regime 1 behaviour (Figure \ref{Regime 1}B) with fewer undulations. The smoother boundary displacement may be attributed to the reduction in the interfacial tensions and the pinning effect from the substrate with ethanol in the displacing liquid. Higher solubility of 1-octanol in ethanol solution may attribute to smaller microdomains.

Consistent results were also observed in Figure \ref{Variations to displacing liquid}C, where solution A was 10\% 1-octanol by mass, and the displacing liquid was 25\% ethanol by mass solution (see E4 in Table \ref{Experimental Conditions Additional Table}). In this case, droplets formed behind the moving boundary and dissolved slowly due to the high solubility of 1-octanol in displacing liquid.

The boundary displacement as a function of time summarized in Figure \ref{Variations to displacing liquid}D shows a time-invariant displacement rate for binary displacing liquid. Conditions with water as the displacing liquid, the boundary displacement rate for Regime 4 is 10 times greater, and for Regime 1, the rate is 0.9 times lower. The faster displacement rate for Regime 1 may be attributed to improvements in solubility of 1-octanol in the displacing liquid.

\subsection{Displacing Water with Solution A}

By displacing water with solution A (a 1-octanol-containing ternary mixture), we are able to consider impacts of reverse flow arising in large scale porous media and to understand the impacts of relative wettability of the displacing liquid. Two cases, linked to experiments described in detail above, are illustrative.

Figure \ref{Variations to displacing liquid}A shows the insoluble boundary arising when the confined liquid is water and the displacing liquid is solution A. The solution A composition corresponds to Regime 1 (see R1 in Table \ref{Experimental Conditions Additional Table}) shown in Figure \ref{Regime 1}A. Droplets did not form within microdomains left behind by the boundary. Instead, small 1-octanol droplets were confined in the boundary zone, similar to Regime 2 conditions in Figure \ref{Regime 2}. The boundary displacement rate was 9 times faster compared to normal Regime 1 conditions.

Figure \ref{Variations to displacing liquid}B shows the displacement boundary when solution A composition corresponds to Regime 4 (see R4 in Table \ref{Experimental Conditions Additional Table}). In this case, 1-octanol-rich droplets and domains form at the boundary . The overall displacement rate of the boundary (Figure \ref{Variations to displacing liquid}C) is time-invariant (77 $\mu m/s$) and 50 times faster than normal Regime 4 conditions. This dramatic difference in displacement rate demonstrates that using displacing liquids with better wettability (lower interfacial tension) on the wall surface may provide an effective route to enhance fluid displacement rate in confinement. Such wettability manipulation is a common professional practice for enhanced oil recovery \cite{Aziz2020}.

\begin{figure}[H]
	\includegraphics[trim={0cm 0cm 0cm 0cm}, clip, width=1\columnwidth]{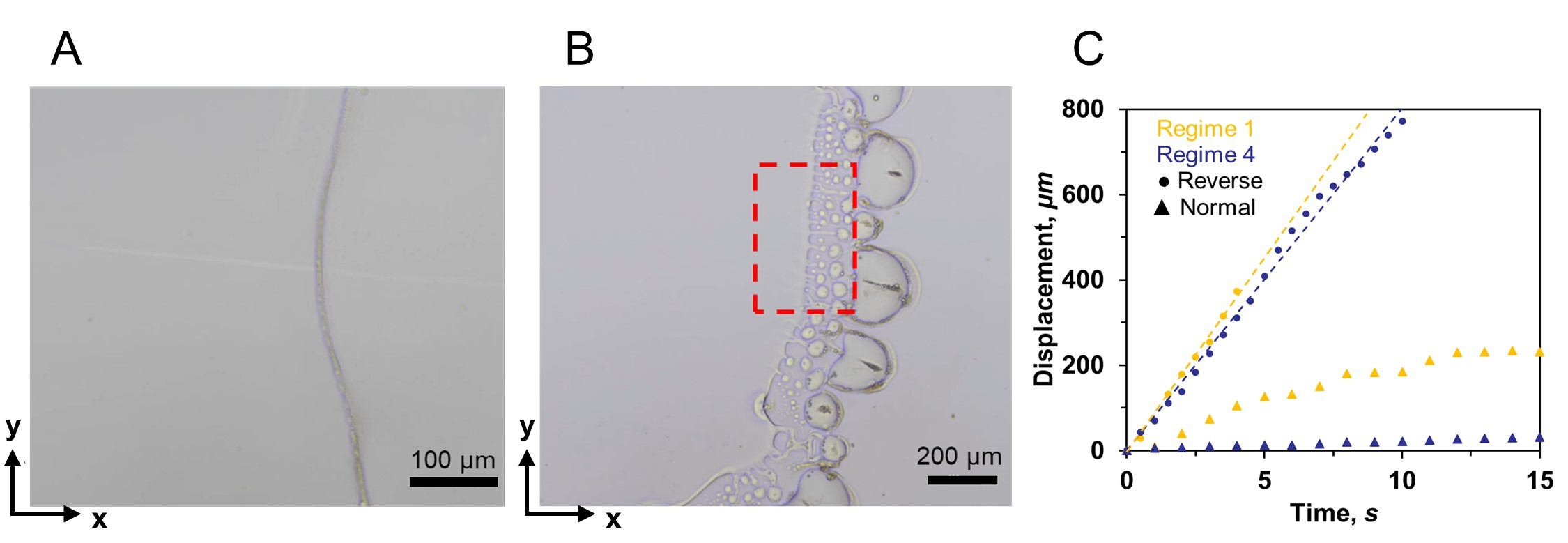}
	\caption{(A) Optical image showing confined water being displaced by solution A with a 50\% 1-octanol by weight composition. (B) Optical image showing confined water being displaced by solution A with a 10\% 1-octanol by weight composition. (C) Boundary displacement as a function time for reverse conditions comparing the 10\% octanol (blue) and 50\% (yellow) conditions. Dashed lines indicate linear regression of data. Data for water as displacing liquid and ternary solution as confined liquid in both regimes is added for comparison (triangles).}
	\label{Variations to displacing liquid 2}
\end{figure}

\section{Conclusions}

We investigate the displacement process for phase-separating ternary liquid mixtures in 2D confinement. The model mixtures contain 1-octanol (a model oil), water (a model poor solvent) and ethanol (a model good solvent), where water (the poor solvent) is the displacing fluid. One of the confining walls is hydrophobic, and the other is hydrophilic. Four composition-dependent displacement regimes are identified. While the details of the displacement mechanism corresponding to each regime differ, order of magnitude enhancements over single-phase diffusive displacement arise. The displacement rate was also shown to be further enhanced with the addition of ethanol (good solvent) to the displacing liquid, improving the wettability on the wall surface. Reverse flow cases where confined water is displaced with the ternary mixture further illustrate the importance of the impact of relative wettability of coexisting phases on boundary displacement. The findings in this work are readily generalized using phase diagrams to provide insights and guidelines for the design of solution formulations for confined liquid displacement. Such displacement processes are important in geological, chemical, and biological processes in enhanced oil recovery, CO$_2$ sequestration, catalytic reactions and drug delivery systems.

\begin{acknowledgement}
This project is supported by the Natural Science and Engineering Research Council of Canada (NSERC) and Future Energy Systems (Canada First Research Excellence Fund). This research was undertaken, in part, thanks to funding from the Canada Research Chairs program.
\end{acknowledgement}

\newpage
\bibliography{literature}

\end{document}


\subsection{Thermodynamic calculations of excessive free energy from phase separation}

\begin{figure}[H]
	\includegraphics[trim={0cm 0cm 0cm 0cm}, clip, width=1\columnwidth]{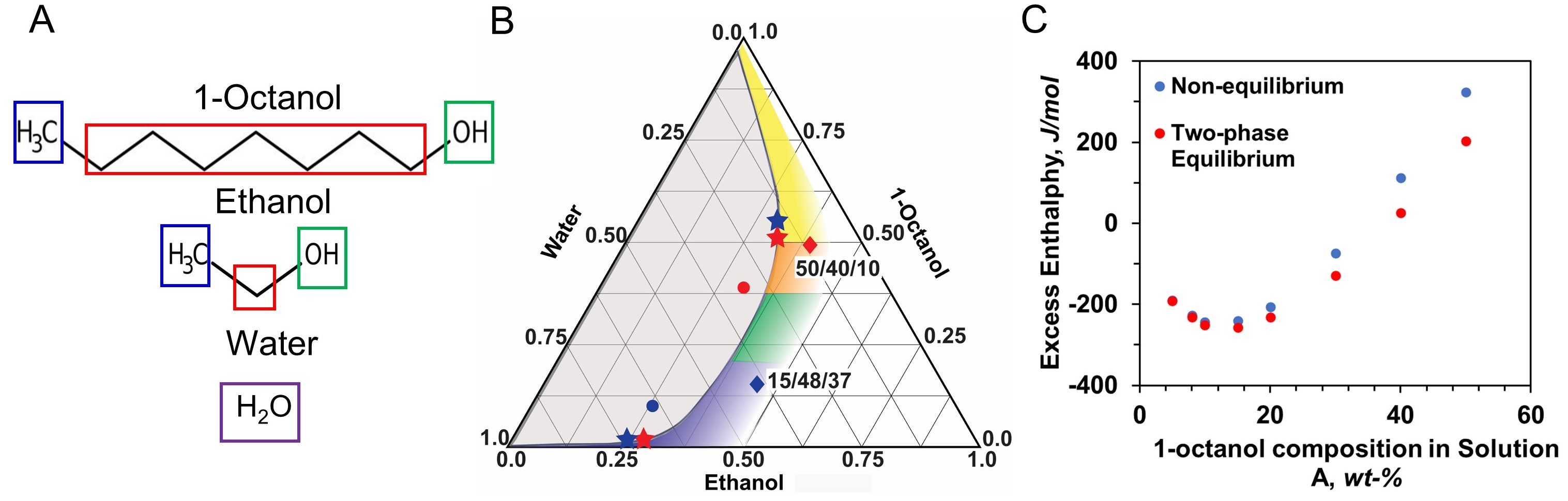}
	\caption{(A) Chemical structure of compounds in system with color coded functional groups used in modified UNIFAC calculations of enthalpy of mixing. Blue (methyl), red (methylene), green (hydroxide), and purple (water). (B) Phase diagram of 1-octanol system showing two of the bulk compositions that were used to calculate the excess enthalpy energy of the different initial compositions one at low 1-octanol concentration (blue) and high 1-octanol concentration (red). Initial compositions (diamond shape), 1-octanol-rich and water-rich subphases (star), and bulk compositions (circles) are all shown in this figure. (C) Calculated excess enthalpy with Dortmund UNIFAC method. Blue data indicates excess enthalpy of mixture at non-equilibrium state. Red data indicates excess enthalpy of mixture at equilibrium state obtained by summation of the enthalpy state of both 1-octanol-rich subphase and water-rich subphase in equilibrium.}
	\label{Thermodynamic Calculations}
\end{figure}

The calculations were performed to approximate the excess energy that was released from the liquid-liquid phase separation. The ternary liquid system was modelled with an activity coefficient model to predict the phase behaviour. The modified UNIFAC model or Dortmund UNIFAC model \cite{Weidlich1987,Gmehling1993} was used as the main thermodynamic model. For the simulations, we utilized both Symmetry Process Simulation software (mass balances) and a MATLAB code (energy balances) based on two works by Fredenslund and Weidlich. The van der Waals properties of the functional groups shown in Figure \ref{Thermodynamic Calculations}A, such as the group volume $R_k$, the surface area $Q_k$ are in Table \ref{vdW UNIFAC}, and the interaction parameters ($a_{mn}$, $b_{mn}$, $c_{mn}$) are to be found in Table \ref{UNIFAC interaction a}, \ref{UNIFAC interaction b} and \ref{UNIFAC interaction c}.

The lowest 1-octanol composition achieved by simulations was the 5\% wt-\% condition by Symmetry. At 5\% condition, we chose the water-rich phase to be around 75 times the amount of the created 1-octanol-rich phase. This was done to have a non-equilibrium composition close to the binodal curve that may approximate the true condition as phase separation occurs in confinement. An example of two initial compositions are depicted in Figure \ref{Thermodynamic Calculations}B. Symmetry is a process design software, meaning every calculation is in terms of a flowrate. To simulate our conditions, we had to fix the initial solution A feed to be a specific mass flowrate quantity. Only solution B was changed for the different initial conditions. The changes in the mass flowrates of B were all decided by a proportionality factor based on an initial optimized value. The proportions are all tabulated with the corresponding phase compositions in Table \ref{sim parameters}. The mass balance simulations were performed using a mixer unit in the program at a temperature of 25 $\degree$C and a pressure of 101.3 kPa.

The compositions from the Symmetry simulations in Table \ref{sim compositions} were then used in the MATLAB code that was created to only calculate the excess enthalpy as a function of the composition of each component. With this, we were thus able to calculate the excess enthalpy of each bulk phase, water-rich and 1-octanol-rich phases. The equilibrium energy states were calculated by summation of each water and 1-octanol-rich phase with their specific fraction in the final state. The non-equilibrium energy states were calculated using the bulk-phase composition. The calculated excess enthalpy values are shown in Figure \ref{Thermodynamic Calculations}C. With these values, it was possible to calculate the energy released for each initial composition, which should provide a positive value for every case.

\begin{table}[H]
\captionsetup{font = {small}}
\caption{Modified UNIFAC van der Waals parameters of functional groups used in simulation for 1-octanol system.}
\centering
\begin{tabular}{c| c c c}
\hline
Functional Group & Classification & Group Volume, & Surface Area, \\
& & $R_k$ & $Q_k$ \\
\hline
Methyl & Main group (n): 1 (-$CH_2$),  & 0.6325 & 1.0608 \\
& Sub-group (m): 1 (-$CH_3$) & & \\
\hline
Methylene & Main group (n): 1 (-$CH_2$),  & 0.6325 & 0.7081 \\
& Sub-group (m): 2 (-$CH_3$) & & \\
\hline
Hydroxide & Main group (n): 5 (-OH),  & 1.2302 & 0.8927 \\
& Sub-group (m): 14 (Primary) & & \\
\hline
Water & Main group (n): 7 (-OH),  & 1.7334 & 2.4561 \\
& Sub-group (m): 16 (Primary) & & \\
\hline
\end{tabular}
\label{vdW UNIFAC}
\end{table}

\begin{table}[H]
\captionsetup{font = {small}}
\caption{UNIFAC modified interaction parameters $a_{mn}$ used in simulation for 1-octanol system.}
\centering
\begin{tabular}{c| c c c c}
\hline
& $CH_{3}$ & $CH_{2}$ & OH & $H_{2}O$ \\
\hline
$CH_{3}$ & 0 & 0 & 2777 & 1391.3 \\
\hline
$CH_{2}$ & 0 & 0 & 2777 & 1391.3 \\
\hline
OH & 1606 & 1606 & 0 & -801.9 \\
\hline
$H_{2}O$ & -17.253 & -17.253 & 1460 & 0 \\
\hline
\end{tabular}
\label{UNIFAC interaction a}
\end{table}

\begin{table}[H]
\captionsetup{font = {small}}
\caption{UNIFAC modified interaction parameters $b_{mn}$ used in simulation for 1-octanol system.}
\centering
\begin{tabular}{c| c c c c}
\hline
& $CH_{3}$ & $CH_{2}$ & OH & $H_{2}O$ \\
\hline
$CH_{3}$ & 0 & 0 & -4.674 & -3.6156 \\
\hline
$CH_{2}$ & 0 & 0 & -4.674 & -3.6156 \\
\hline
OH & -4.746 & -4.746 & 0 & 3.824 \\
\hline
$H_{2}O$ & 0.8389 & 0.8389 & -8.673 & 0 \\
\hline
\end{tabular}
\label{UNIFAC interaction b}
\end{table}

\begin{table}[H]
\captionsetup{font = {small}}
\caption{UNIFAC modified interaction parameters $c_{mn}$ used in simulation for 1-octanol system.}
\centering
\begin{tabular}{c| c c c c}
\hline
& $CH_{3}$ & $CH_{2}$ & OH & $H_{2}O$ \\
\hline
$CH_{3}$ & 0 & 0 & 0.001551 & 0.001144 \\
\hline
$CH_{2}$ & 0 & 0 & 0.001551 & 0.001144 \\
\hline
OH & 0.0009181 & 0.0009181 & 0 & -0.007514 \\
\hline
$H_{2}O$ & 0.0009021 & 0.0009021 & 0.01641 & 0 \\
\hline
\end{tabular}
\label{UNIFAC interaction c}
\end{table}

\begin{table}[H]
\captionsetup{font = {small}}
\caption{The multiplier factors were used with the initial flowrate of the 5 \% 1-octanol condition as basis, which was found to be 1.9 $m^{3}/h$ from optimization with a fixed Solution A flowrate of 1 $m^{3}/h$. The optimization was based on having the set initial condition of 75 times the mass amount of water-phase to 1-octanol-rich phase.}
\centering
\begin{tabular}{c c c c}
\hline
Condition & Multiplier & Solution A & Solution B \\
& & $m^3/h$ & $m^3/h$\\
\hline
5/48/47 & 1 & 1.00 & 1.90 \\
\hline
8/47/45 & 0.625 & 1.00 & 1.19 \\
\hline
10/48/42 & 0.5 & 1.00 & 0.95 \\
\hline
15/48/37 & 0.333 & 1.00 &  0.63 \\
\hline
20/48/32 & 0.25 & 1.00 & 0.48 \\
\hline
30/48/22 & 0.167 & 1.00 & 0.32 \\
\hline
40/45/15 & 0.125 & 1.00 & 0.24 \\
\hline
50/40/10 & 0.1 & 1.00 & 0.19 \\
\hline
\end{tabular}
\label{sim parameters}
\end{table}

\begin{table}[H]
\captionsetup{font = {small}}
\caption{Compositions obtained from Symmetry for simulations described in terms of mass percentage 1-octanol/ethanol/water.}
\centering
\begin{tabular}{c c c c c}
\hline
Condition & Bulk Phase & Water-rich Phase & 1-Octanol-rich Phase & W/O Fraction \\
\hline
5/48/47 & 1.58/15.18/83.24 & 0.64/15.12/84.23 & 71.37/19.11/9.52 & 98.7/1.3 \\
\hline
8/47/45 & 3.40/19.95/76.66 & 0.94/19.76/79.29 & 64.16/24.47/11.37 & 96.12/3.88 \\
\hline
10/48/42 & 4.78/22.95/72.27 & 1.19/22.64/76.16 & 59.72/27.63/12.65 & 93.87/6.13 \\
\hline
15/48/37 & 8.65/27.68/63.67 & 1.71/26.99/71.31 & 53.08/32.13/14.79 & 86.48/13.52 \\
\hline
20/48/32 & 12.85/30.85/56.29 & 2.14/17.20/80.26 & 48.95/34.77/16.28 & 77.1/22.90 \\
\hline
30/48/22 & 21.78/34.85/43.36 & 2.74/32.63/64.63 & 44.43/37.50/18.07 & 54.33/45.67 \\
\hline
40/45/15 & 31.10/34.99/33.91 & 2.54/31.73/65.73 & 45.84/36.67/17.49 & 34.03/65.97 \\
\hline
50/40/10 & 40.64/32.51/26.85 & 1.93/28.46/69.61 & 50.85/33.58/15.57 & 20.86/79.14 \\
\hline
\end{tabular}
\label{sim compositions}
\end{table}

\section{Supporting figures}

\begin{figure}[H]
	\includegraphics[trim={0cm 0cm 0cm 0cm}, clip, width=0.95\columnwidth]{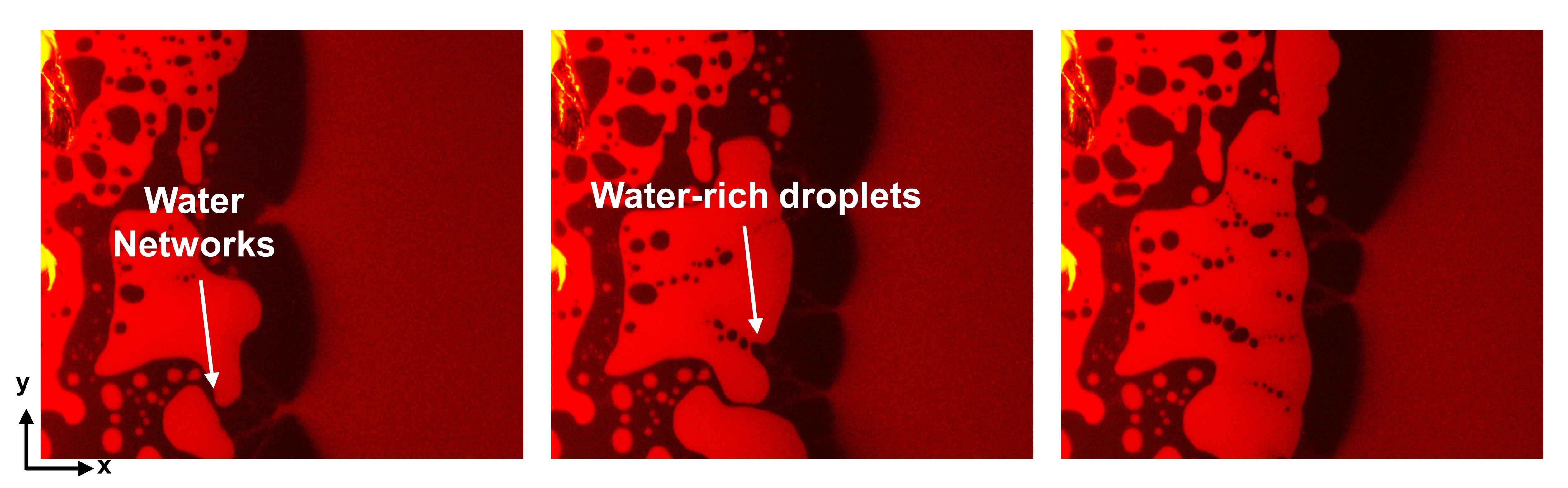}
	\caption{Optical images showing early development of the water-rich zone 2 within Regime 3 conditions.}
	\label{supporting}
\end{figure}

\section{Supporting videos}

\noindent Video S1: Bright-field imagery of boundary displacement at Regime 1 (50\% 1-octanol by mass in Solution A) condition. Video shows boundary undulation and microdomain formation.

\noindent Video S2: Bright-field imagery of boundary displacement at Regime 2 (40\% 1-octanol by mass in Solution A) condition. Video shows boundary undulation and stagnant droplets left behind by the boundary.

\noindent Video S3: Boundary displacement at Regime 3 (30\% 1-octanol by mass in Solution A) condition. Video from 0 to 8 seconds shows the fluorescence imagery of the three-zone configuration at the boundary. From 8 to 20 seconds the fluorescence mode was changed to bright-field imaging.

\noindent Video S4: Boundary displacement at Regime 4 (10\% 1-octanol by mass in Solution A) condition. Video showing the transition from a triangular protrusion in the boundary to a line shaped protrusion. From 0 to 5 the video was in fluorescence mode, and from 5 to 25 it was in bright-field mode.

\newpage
\bibliography{literature}